# Cryptanalysis of Yang-Wang-Chang's Password Authentication Scheme with Smart Cards

Al-Sakib Khan Pathan and Choong Seon Hong
Department of Computer Engineering, Kyung Hee University, Korea
spathan@networking.khu.ac.kr and cshong@khu.ac.kr

*Abstract* — In 2005, Yang, Wang, and Chang proposed an improved timestamp-based password authentication scheme in an attempt to overcome the flaws of Yang-Shieh's legendary timestamp-based remote authentication scheme using smart cards. After analyzing the improved scheme proposed by Yang-Wang-Chang, we have found that their scheme is still insecure and vulnerable to four types of forgery attacks. Hence, in this paper, we prove that, their claim that their scheme is intractable is incorrect. Also, we show that even an attack based on Sun et al.'s attack could be launched against their scheme which they claimed to resolve with their proposal.

## 1. Introduction

Remote user authentication using smart cards is a well-known method used in many electronic networking applications like e-trade, e-transactions, e-banking etc. With the growth of Internet and different networking applications, it has become a popular method for secure remote communications. So far, a number of timestamp-based remote user authentication schemes have been proposed, many of which cannot sustain to certain types of forgery attacks. In 1999, Yang and Shieh [1] proposed two authentication schemes using smart cards, one of which was the time-stamp based password authentication scheme. Later this scheme was found as faulty by several researchers and also some modifications were proposed. Yang-Wang-Chang's scheme [2] has so far been considered as one of the well-standing alternate improved solutions for overcoming the vulnerabilities of [1]. But in this paper, we show that, their scheme is still vulnerable to four different types of forgery attacks, in spite of their claim of the scheme's intractability.

The main contributions of this paper are:

1. We review the time-stamp based authentication scheme proposed by Yang, Wang, and Chang.

2. We show four different types of forgery attacks that could be launched against Yang-Wang-Change scheme and thus prove that, their scheme is not fully secure as they claimed in their paper.

The rest of the paper is structured as follows; Section 2 states the motivation for our work, Section 3 reviews the working method of Yang-Wang-Chang scheme, Section 4 presents the cryptanalysis and weaknesses of Yang-Wang-Chang scheme, and finally Section 5 concludes the paper summarizing our contributions with future research directions.

## 2. Motivation of Our Work

In an attempt of removing the flaws of Yang-Shieh's scheme [1], Yang, Wang, and Chang have improved the working principle of [1] with some modifications in the computations of some parameters and in some checking steps. They, in their paper have referred to the attacks on [1] devised by Chan and Cheng [3]. They have also pointed how Sun and Yeh [4] showed the vulnerability of [1]. Hence, they went for an improvement for the scheme presented in [1].

After analyzing the improved scheme proposed by Yang-Wang-Chang, we have found that, it is still vulnerable to a type of forgery attack based on [4]. Also, we devise other new forgery attacks that could be applied against their scheme. Hence, we prove that, Yang-Wang-Chang scheme could not be a secure scheme to be used for password authentication of the remote users using smart cards.

## 3. Review of Yang-Wang-Chang Scheme

In this section, we review Yang-Wang-Chang password authentication scheme [2] using smart cards. First, we note down the terms and preliminaries used throughout the rest of the paper.

### 3.1 Basic Terms

$U_i$ – The *i*th user seeking for authentication by the server
KIC – The Key Information Center which is responsible for generating key information, issuing smart cards to new users, and serving password-changing requests for registered users
$ID_i$ – The identity of the user $U_i$
$PW_i$ – The password chosen by the user $U_i$
$CID_i$ – The identity of the smart card generated by the server associated with the user $U_i$
$f(.)$ – A one-way function

This work was supported by ITRC and MIC. Dr. CS Hong is the corresponding author.

## 3.2 Yang-Wang-Chang's Timestamp-Based Scheme

As this scheme is a modified and enhanced version of Yang-Shieh scheme [1], it also has three phases for the password authentication process; registration phase, login phase, and authentication phase. The three phases are described in the following subsections.

**Registration Phase.** In this phase, the KIC sets up the authentication system and issues smart cards to user $U_i$ who requests for registration. It is assumed that, this phase occurs over a secure channel. At first, the $U_i$ securely submits $ID_i$ and $PW_i$ to the KIC. Then, the steps that the KIC follows are:

1. Two large prime numbers $p$ and $q$ are generated, and $n = p \cdot q$ is computed.
2. A public key $e$ and corresponding secret key $d$ are chosen which satisfy, $e \cdot d \equiv 1 \mod (p-1)(q-1)$.
3. An integer $g$ is found which is a primitive element in both GF($p$) and GF($q$), where $g$ is the public information of the KIC. It should be noted that, GF($p$) and GF($q$) mean $p$ and $q$ are finite fields.
4. A smart card identifier $CID_i$ is generated for the user $U_i$ and $S_i$ is computed as, $S_i = ID_i^{CID_i \cdot d} \mod n$.
5. $h_i$ for $U_i$ is computed such that, $h_i = g^{PW_i \cdot d} \mod n$.
6. Then the information $n$, $e$, $g$, $ID_i$, $CID_i$, $S_i$, $h_i$ and $f(.)$ are written into the smart card's memory and the card is issued to the user $U_i$

**Login Phase.** When $U_i$ needs to login to the system, the smart card should be attached to the login device, and $ID_i$ and $PW_i$ need to be keyed in. After that, the smart card performs the following operations:

1. Generates a random number $r_i$
2. Computes $X_i$ and $Y_i$ as follows:

$$X_i = g^{PW_i \cdot r_i} \mod n$$
$$Y_i = S_i \cdot h_i^{r_i \cdot T} \mod n$$

where, T is the current timestamp.

3. Sends the login request message, $M = \{ID_i, CID_i, X_i, Y_i, n, e, g, T\}$ to the remote server.

**Verification Phase.** In this phase, the system or the server determines the validity of the received login request message and decides whether to accept the access of the user or not. So, after the server has received the message $M$, it carries out the following steps:

1. Checks the validity of $ID_i$ and $CID_i$. If the formats of these are incorrect, the server rejects the login request.
2. Checks whether the condition $(T'-T) \leq \Delta T$ holds or not, where $T'$ is the timestamp of receiving the login request message and $\Delta T$ is the legitimate time interval allowed for the transmission delay. If it is negative, the server rejects the request.

3. Checks the equation, $Y_i^e \equiv ID_i^{CID_i} \cdot X_i^T \mod n$. If it holds, then the remote server accepts the login request and gives access to the $U_i$.

## 4. Forgery Attacks on Yang-Wang-Chang's Scheme

In this section, we show the vulnerabilities of Yang-Wang-Chang's proposed scheme. We show first two attacks based on some of the previously shown attacks on timestamp based password authentication scheme. Third and fourth attacks are our novel attacks that could also be launched successfully on Yang-Wang-Chang's scheme.

### 4.1 Attack Based on [4] and [5]

As the attacker could intercept the login request message $M = \{ID_i, CID_i, X_i, Y_i, n, e, g, T\}$, it can get the valid values of $ID_i$ and $CID_i$. Using these values it could launch a forgery attack as follows:

1. Let $T_a$ be the timestamp for the attacker's login request. Use the Extended Euclidean algorithm to compute $gcd(e, T_a) = 1$ that is, e and $T_a$ are relatively prime. Let, $u$ and $v$ be the coefficients computed by the extended Euclidean algorithm such that, $e \cdot u - T_a \cdot v = 1$
2. Compute $X_f = ID_i^{CID_i \cdot v} \mod n$
3. Compute $Y_f = ID_i^{CID_i \cdot u} \mod n$
4. Send the forged login request message $M_f = \{ID_i, CID_i, X_f, Y_f, n, e, g, T_a\}$ and this request will eventually pass the authentication phase as,

$$Y_f^e = (ID_i^{CID_i \cdot u})^e \mod n$$
$$= (ID_i^{CID_i})^{eu} \mod n$$
$$= (ID_i^{CID_i})^{1+T_a \cdot v} \mod n$$
$$= ID_i^{CID_i} \cdot (ID_i^{CID_i})^{T_a \cdot v} \mod n$$
$$= ID_i^{CID_i} \cdot (X_f)^{T_a} \mod n$$

In fact, this attack could be extended for $gcd(e,a)=2,3,\ldots$ instead of only $gcd(e, a) = 1$. Hence, the claim that their improved scheme is resistant to Sun et al.'s [4] attack or similar attacks is not correct.

### 4.2 Forgery Attack Based on Yang et al.'s [6] Attack

In this attack the attacker intercepts the message, $M = \{ID_i, CID_i, X_i, Y_i, n, e, g, T\}$ and then:

1. Finds a value $w$ such that it satisfies, $w \cdot T_a = T$, where $T_a$ denotes the attacker's attack launching time.
2. Computes the equation, $X_f = X_i^w = g^{r_i \cdot PW_i \cdot w} \mod n$
3. Now, the attacker constructs the forged login request message as, $M_f = \{ID_i, CID_i, X_f, Y_i, n, e, g, T_a\}$

This forged message eventually passes the authentication phase of [2] because:

$$Y_i^e = (S_i \cdot h_i^{r_i \cdot T})^e \bmod n$$
$$= (ID_i^{CID_i \cdot d} \cdot g^{PW_i \cdot d \cdot r_i \cdot T})^e \bmod n$$
$$= ID_i^{CID_i} \cdot g^{PW_i \cdot r_i \cdot T} \bmod n$$

and,
$$ID_i^{CID_i} \cdot X_f^{T_a} \bmod n = ID_i^{CID_i} \cdot g^{r_i \cdot PW_i \cdot w \cdot T_a} \bmod n$$
$$= ID_i^{CID_i} \cdot g^{PW_i \cdot r_i \cdot T} \bmod n$$

### 4.3 Our Novel Impersonation Attack

An attacker can impersonate a legitimate user $U_i$, with identity $ID_i$, by using the following procedure:

1. It Intercepts the login request message $M = \{ID_i, CID_i, X_i, Y_i, n, e, g, T\}$.
2. Computes, $ID_f = ID_i^{-1} \bmod n$
3. Now, the attacker submits the identity $ID_f$ and a random value as his password to the KIC to obtain a valid smart card with information $\{n, e, g, ID_f, CID_f, S_k, h_k \text{ and } f(.)\}$.
4. Since, in the registration phase, $S_i = ID_i^{CID_i \cdot d} \bmod n$ and here, $S_k = ID_f^{CID_i \cdot d} \bmod n = ID_i^{-CID_i \cdot d} \bmod n$, the attacker can compute $S_i$ as, $S_i = S_k^{-1} \bmod n$
5. Then, the attacker chooses a random integer $y$.
6. Sets, $X_f = y^e \bmod n$ and $Y_f = S_i \cdot y^{T_f} \bmod n$, where $T_f$ is the timestamp for the login request from the attacker and sends the forged login message, $M_f = \{ID_i, CID_i, X_f, Y_f, n, e, g, T_f\}$. The request is validated as the login request from the user $U_i$ because,

$$Y_f^e = (S_i \cdot y^{T_f})^e \bmod n$$
$$= (S_k^{-1} \cdot y^{T_f})^e \bmod n$$
$$= ID_i^{CID_i \cdot d \cdot e} \cdot y^{T_f \cdot e} \bmod n$$
$$= ID_i^{CID_i} \cdot (X_f)^{T_f} \bmod n$$

### 4.4 Our Novel Forgery Attack

The attacker can get the values of $ID_i$ and $CID_i$ from the login request message from the valid user, and the smart card identifier $CID_i$ is a fixed value for a particular login request from a user. The attacker could launch an attack using the following steps:

1. The attacker finds a value $T_f$ such that, $T \cdot T_f \equiv 1 \bmod n$ where $T_f$ is the attacker's login timestamp
2. It chooses a random integer $k$ and computes, $Y_f = k^{T_f} \bmod n$ and sets $X_f = ID_i^{-CID_i \cdot T} \cdot k^e \bmod n$.
3. Sends the forged login request message, $M_f = \{ID_i, CID_i, X_f, Y_f, n, e, g, T_f\}$
4. The attacker could pass the authentication phase as,

$$Y_f^e = k^{e \cdot T_f} \bmod n$$

and,
$$ID_i^{CID_i} \cdot X_f^{T_f} \bmod n = ID_i^{CID_i} \cdot (ID_i^{-CID_i \cdot T} \cdot k^e)^{T_f} \bmod n$$
$$= ID_i^{CID_i} \cdot ID_i^{-CID_i \cdot T \cdot T_f} \cdot k^{e \cdot T_f} \bmod n$$
$$= ID_i^{CID_i} \cdot ID_i^{-CID_i} \cdot k^{e \cdot T_f} \bmod n$$
$$= k^{e \cdot T_f} \bmod n$$

## 5. Conclusions and Future Works

After analyzing the Yang-Wang-Change scheme, we have found that, it fails to pass some tricky impersonation or forgery attacks. In this paper, with four different types of forgery attacks, we show that Yang-Wang-Change password authentication scheme is still vulnerable and hence, it cannot provide proper security according to their claim. As our future work, we would like to find out a solution for fixing up the vulnerabilities and to propose an improved scheme that could resist all these forgery attacks.

REFERENCES

[1] Yang, W.-H. and Shieh, S.-P., "Password Authentication Schemes with Smart Cards", Computers & Security, Vol. 18, No. 8, Elsevier, 1999, pp. 727-733.
[2] Yang, C. C., Wang, R.-C., and Chang, T.-Y., "An improvement of the Yang-Shieh password authentication schemes", Applied Mathematics and Computation 162, Elsevier, 2005, pp. 1391-1396.
[3] Chan, C.-K. and Cheng, L. M., "Cryptanalysis of a Timestamp-Based Password Authentication Scheme", Computers and Security, Vol. 21, No. 1, Elsevier, 2002, pp. 74-76.
[4] Sun, H.-M. and Yeh, H.-T., "Further Cryptanalysis of a Password Authentication Scheme with Smart Cards", IEICE Transactions on Communications, Vol. E86-B, No. 4, 2003, pp. 1212-1215.
[5] Chen, K.-F. and Zhong, S., "Attacks on the (Enhanced) Yang-Shieh Authentication", Computers & Security, Vol. 22, No. 8, Elsevier, 2003, pp. 725-727.
[6] Yang, C.-C., Yang, H.-W., and Wang, R. C., "Cryptanalysis of Security Enhancement for the Timestamp-Based Password Authentication Scheme using Smart Cards", IEEE Transactions on Consumer Electronics, Vol. 50, No. 2, 2004, pp. 578-579.